\documentclass[3p,times,procedia]{elsarticle}
\usepackage{nupha_ecrc}

\volume{00}

\firstpage{1}

\journalname{Nuclear Physics A}

\runauth{}

\jid{nupha}

\jnltitlelogo{Nuclear Physics A}

\usepackage{amssymb}
\usepackage{amsmath}
\usepackage[figuresright]{rotating}

\begin{document}

\begin{frontmatter}



%% Instructions from Editor: Please use the following \dochead only in the preprint version (e-print arXiv etc.); 
%% use empty \dochead{} when submitting to Nuclear Physics A!
\dochead{XXVIIIth International Conference on Ultrarelativistic Nucleus-Nucleus Collisions\\ (Quark Matter 2019)}
%% e.g. \dochead{17th International Conference on Dynamical Processes in Excited States of Solids}

\title{Exploring Longitudinal Observables with 3+1D IP-Glasma}


\author{Scott McDonald, Sangyong Jeon, Charles Gale}

 \address{Department of Physics, McGill University, 3600 University
 Street, Montreal,
 QC, H3A 2T8, Canada}

\begin{abstract}
We present a formulation of the initial state of heavy ion collisions that generalizes the 2+1D boost invariant IP-Glasma \cite{Schenke:2012wb} to 3+1D through JIMWLK rapidity evolution of the pre-collision Wilson lines.  The rapidity dependence introduced by the JIMWLK evolution leads us to modify the initial condition for the gauge fields, and to solve Gauss' law iteratively in order to allow for temporal evolution on a 3-dimensional lattice.

While the transverse physics of QGP has been studied nearly exhaustively, the effect of longitudinal fluctuations introduced by the JIMWLK evolution has yet to be studied in detail phenomenologically. Hence, we couple our 3+1D IP-Glasma model to MUSIC+UrQMD, for completely 3+1D simulations of heavy ion collisions. Specifically, we consider Pb-Pb collisions at $\sqrt{s} = 2.76\, {\rm TeV}$ and study the rapidity dependence of the charged hadron $v_n(\eta)$ via the $\eta$-dependent flow factorization ratios $r_n(\eta_a,\eta_b)$ as measured by CMS \cite{Khachatryan:2015oea}, as well as the charged hadron multiplicity $dN_{ch}/d\eta$.
\end{abstract}


\end{frontmatter}

% \linenumbers

\section{Introduction}
Quark Gluon Plasma, a thermalized state of deconfined quarks and gluons, is believed to be formed in Heavy Ion Collisons (HIC's) conducted at RHIC and the LHC. Utilizing the assumption of boost invariance, the transverse structure of HIC's has been studied in great detail. In this work, we relax the assumption of boost invariance to investigate the phenomenological effects of realistic initial state rapidity fluctuations arising from JIMWLK evolution of the pre-collision Wilson lines in the 3+1D IP-Glasma model. Coupling this to MUSIC+UrQMD, we investigate the rapidity structure of HIC's.

\section{Model}
The inclusion of the JIMWLK rapidity evolution to generalize the IP-Glasma model to 3+1D was first done in \cite{Schenke:2016ksl}. Numerically, the form of the JIMWLK equation used here is the Langevin formulation from \cite{Lappi:2012vw} which evolves the Wilson lines $V_{A,B}(\mathbf{x}, Y)$ in rapidity,
\begin{equation}\label{eq:LangevinStep}
\begin{split}
     V_{A,B}(\mathbf{x},& Y+dY)=\exp\left({-i\frac{\sqrt{ dY}}{\pi}\int_{\mathbf{u}} \mathbf{K}_{\mathbf{x}-\mathbf{u}} (\cdot V_u\mathbf{\eta_u}V_u^\dagger)}\right) V_{A,B}(\mathbf{x}, Y) \exp \left({i\frac{\sqrt{dY}}{\pi}\int_{\mathbf{v}} \mathbf{K}_{\mathbf{x}-\mathbf{v}} \cdot \mathbf{\eta_v}}\right)
\end{split} 
\end{equation}
 where A(B) label the two nuclei and $\mathbf{\eta}=\left( \eta_1^a t^a,\eta_2^a t^a \right)$. The correlator for the noise term, $\mathbf{\eta}$, is given by,
\begin{equation}
     \langle \mathbf{\eta}_{\mathbf{x}}^{a,i}\mathbf{\eta}_{\mathbf{y}}^{b,j}\rangle=\delta^{ab}\delta^{ij}\int \frac{d^2\mathbf{k}}{(2\pi)^2}e^{i\mathbf{k} \cdot (\mathbf{x-y})} \alpha_s(\mathbf{k}).
  \end{equation}
where the running coupling prescription from \cite{Lappi:2012vw} is used. The modified kernel, as used in \cite{Schenke:2016ksl}, is given by,
\begin{align}
K_{\mathbf{x-z}}&=m|\mathbf{x}-\mathbf{z}|K_1(m|\mathbf{x}-\mathbf{z}|)\frac{\mathbf{x-z}}{(\mathbf{x-z})^2}
\end{align}
where $K_1(x)$ is the Bessel function of the second kind. 

The JIMWLK evolution provides realistic longitudinal fluctuations, while the typical CGC initial condition assumes boost invariance, i.e. vanishing longitudinal derivatives. The rapidity dependence introduced by the JIMWLK evolution means that 1) the form of the 2+1D initial gauge fields will give rise to non-physical energy deposition outside of the interaction region of the colliding nuclei and 2) the system no longer trivially satisfies Gauss’ law at the initial time. In \cite{McDonald:2018wql}, the typical 2D initial condition for the gauge fields was altered from pure gauge in the transverse directions to pure gauge in all three spatial dimensions,
%==========================================%
\begin{align}\label{eq:Aeta}
      A_{i=x,y,\eta}=A^{A}_{i}+A^{B}_{i} && A^{A(B)}_{i=x,y,\eta} = \frac{i}{g}V^{A(B)} \partial_i V^{A(B)\dagger}
\end{align}
%==========================================%
%%%%%%%%%%%%%%%%%%%%%%%%%%%%%%
\begin{figure}[h!]
    \centering
    \includegraphics[width=0.7\linewidth]{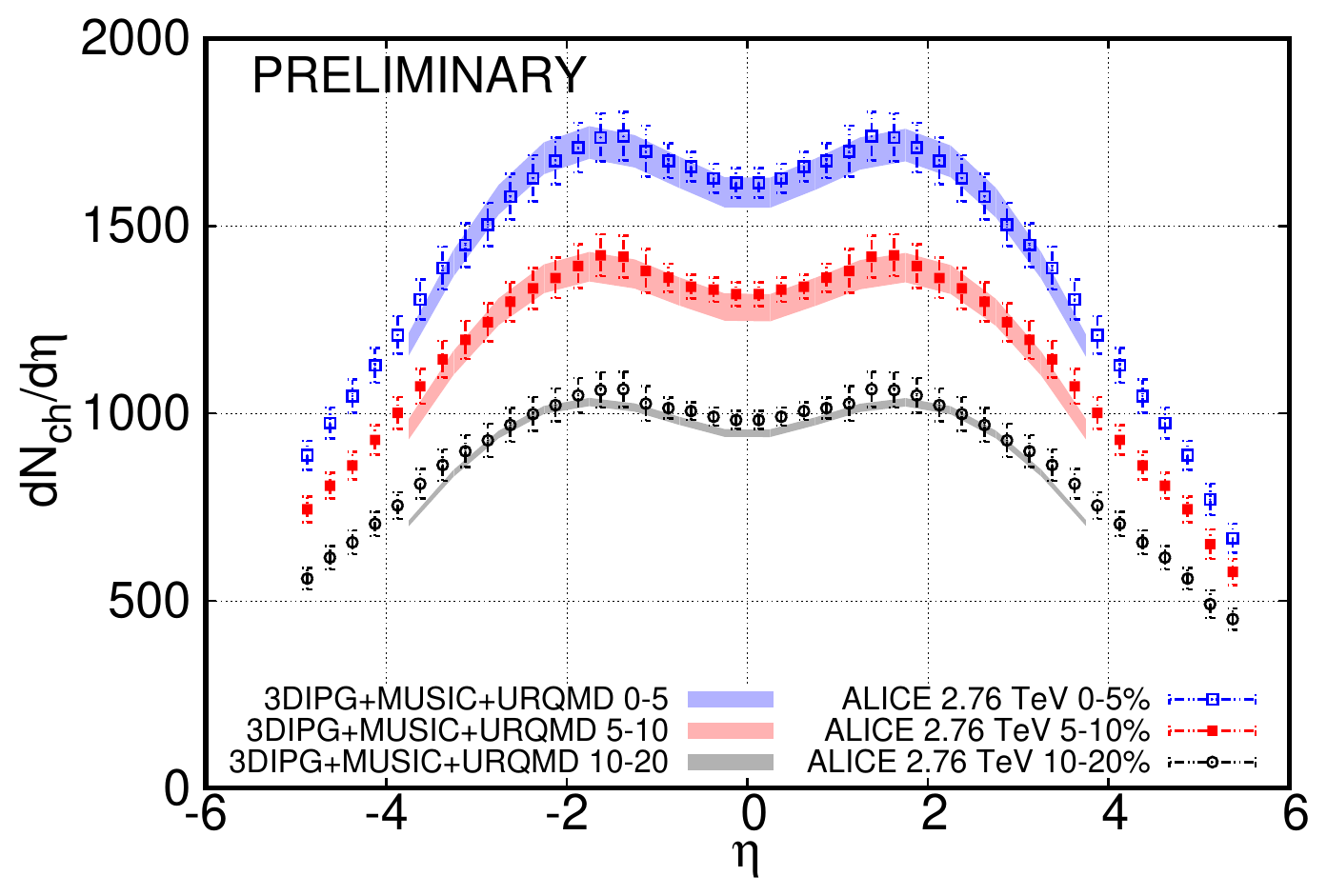}
    \caption{Data for Pb+Pb collisions at 2.76 TeV from the ALICE collaboration \cite{Abbas:2013bpa}. Note that part of the rapidity structure in the calculation is put by hand to avoid numerical problems (see Eq. (\ref{eq:eta_profile})).}
    \label{fig:dNdy}
\end{figure}
%%%%%%%%%%%%%%%%%%%%%%%%%%%%%%
where A(B) label the two pre-collision nuclei. This takes care of the non-physical energy deposition outside of the interaction region. Additonally, the following ansatz for the initial transverse electric fields was taken
%==========================================%
\begin{align}
    E^i=[D^i,\phi]  &&   [D_\eta, E^\eta]+[D_i,E^i]=[D_\eta, E^\eta]+[D_i,[D^i,\phi]]=0.
\end{align}
%==========================================%
which turns Gauss' Law into the covariant Poisson equation. These changes to the initial condition allow for self-consistent temporal evolution of the the Classical Yang Mills equations on a 3-dimensional lattice, and thus for phenomenological application with 3+1D hydrodynamic evolution. 

The initial state events are matched to MUSIC \cite{Schenke:2010nt} at $\tau_0=0.6 \,{\rm fm}$ with a rapidity profile that does not alter the initial condition for $|\eta|<2.5$, but that smoothly brings the energy density to zero in the region $|\eta|>2.5$, to avoid numerical problems from derivatives in the $\eta$-direction or boundary effects. The profile is of the form,
\begin{equation}\label{eq:eta_profile}
    T^{\mu\nu}_{{\rm hydro}}(\eta, \tau_0) = T^{\mu\nu}_{{\rm IP-Glasma}}(\eta, \tau_0)\theta\left(|\eta|-2.5|\right)\exp{\left[-\frac{(|\eta|-2.5)^2}{2}\right]}.
\end{equation}
MUSIC is evolved in 3+1D with a temperature dependent bulk viscosity $\zeta/s(T)$ and $\eta/s=0.10$. It is evolved to a constant temperature hypersurface of $T_{sw}=145 \, {\rm MeV}$, at which point UrQMD \cite{Bass:1998ca} is used for the hadronic gas phase.  

\section{Results}
\begin{figure}[h!]
    \centering
    \includegraphics[width=1.0\textwidth]{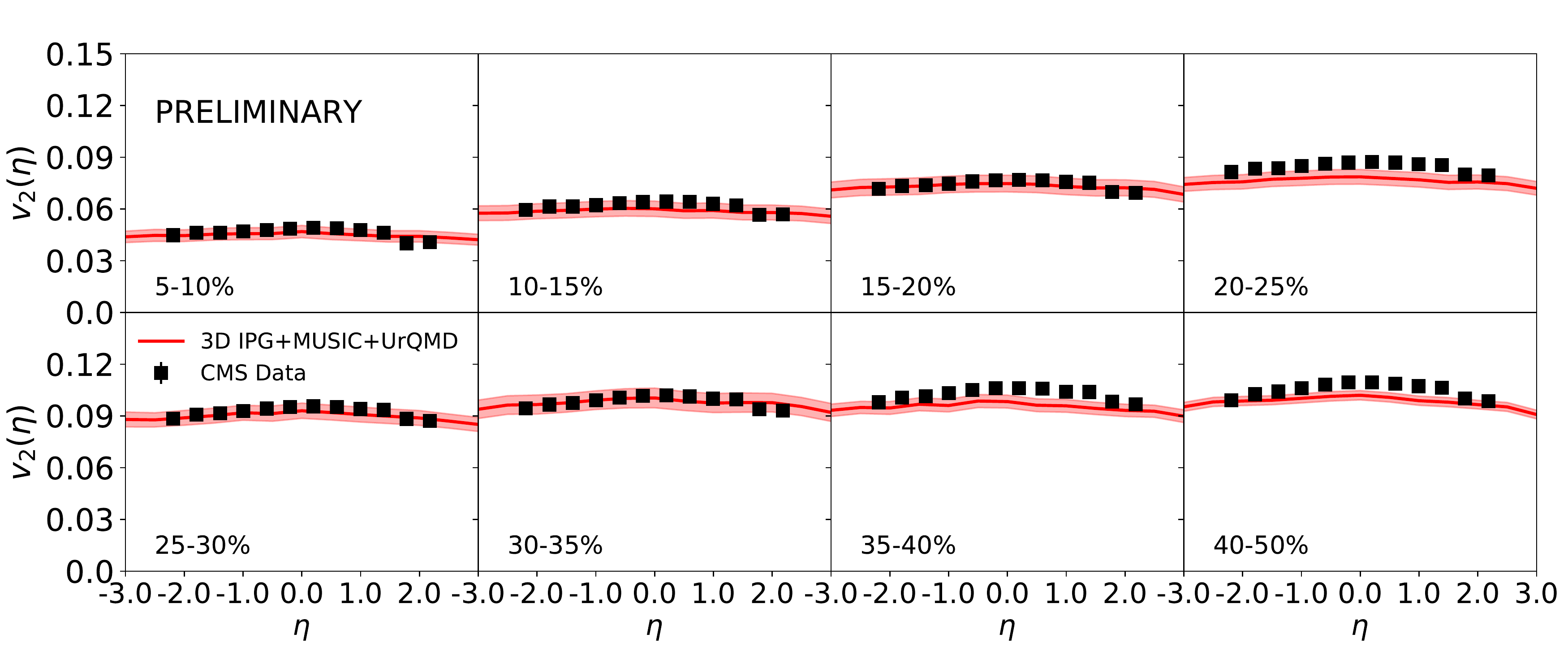}
    \caption{Model calculation for $v_2(\eta)$ compared to CMS data from \cite{Chatrchyan:2012ta} for eight centrality classes.}
    \label{fig:v2}
\end{figure}

With a 3+1D initial condition and hydrodynamic evolution, it is possible to study the full dynamics of heavy ion collisions, and it is particularly interesting to explore the longitudinal structure. In Fig. (\ref{fig:dNdy}), the pseudo-rapidity dependence of charged hadron multiplicity dependence is shown, compared to ALICE \cite{Abbas:2013bpa} data for three centralities, and good agreement is achieved. In Fig. (\ref{fig:v2}), the rapidity dependence of $v_2$ is shown for eight centrality classes. The average $v_2$ is nearly flat in the rapidity range shown, but the magnitude and rapidity dependence of the model agrees quite well with CMS data \cite{Chatrchyan:2012ta}. Transport coefficients are tuned to CMS data for $v_2$ in order to optimize comparison with CMS data for $r_n(\eta_a, \eta_b)$. 
%%%%%%%%%%%%%%%%%%%%%%
%%%%%%%%%%%%%%%%%%%%%%
The rapidity fluctuations introduced by the JIMWLK evolution break boost invariance and lead to rapidity decorrelation in the event-planes and magnitudes of $v_2$ and $v_3$. Here, we compare the correlation in the initial state anisotropies and the final state flow correlations after hydrodynamic evolution,
%%%%%%%%%%%%%%%%%%%%%%%%%%%%%%%%%%%%%%%%%%%
\begin{align}
        r_n(\eta_a, \eta_b)= \frac{\langle v_n(-\eta_a)v_n(\eta_b)\cos(n(\phi_n(-\eta_a)-\phi(\eta_b)) \rangle}{\langle v_n(\eta_a)v_n(\eta_b)\cos(n(\phi_n(\eta_a)-\phi(\eta_b)) \rangle} &&
        \tilde{r}_n(\eta_a, \eta_b)= \frac{\langle \varepsilon_n(-\eta_a)\varepsilon_n(\eta_b)\cos(n(\psi_n(-\eta_a)-\psi(\eta_b)) \rangle}{\langle \varepsilon_n(\eta_a)\varepsilon_n(\eta_b)\cos(n(\psi_n(\eta_a)-\psi(\eta_b)) \rangle}
\end{align}
%%%%%%%%%%%%%%%%%%%%%%%%%%%%%%%%%%%%%%%%
where $\eta_b=3.5$.  Here,  $v_n$ is the n$^{\rm th}$ order flow harmonic and $\varepsilon_n$ is the n$^{\rm th}$ order initial state energy anisotropy,
%%%%%%%%%%%%%%%%%
\begin{equation}
  \mathbf{\varepsilon_n}(\eta)= \frac{\int d^2x r^n \epsilon(\mathbf{x_\perp}, \eta) e^{in\phi}}{\int d^2x r^n \epsilon(\mathbf{x_\perp}, \eta)}.
\end{equation}
%%%%%%%%%%%%%%%%%
Typically, in computing the eccentricities, one re-centers around the center of energy density in the transverse plane. To avoid having different centers for the the different rapidity slices, all rapidity slices are centered around the center of energy density at mid-rapidity ($\eta=0)$.
\begin{figure}[h!]
    \centering
    \includegraphics[width=1.0\linewidth]{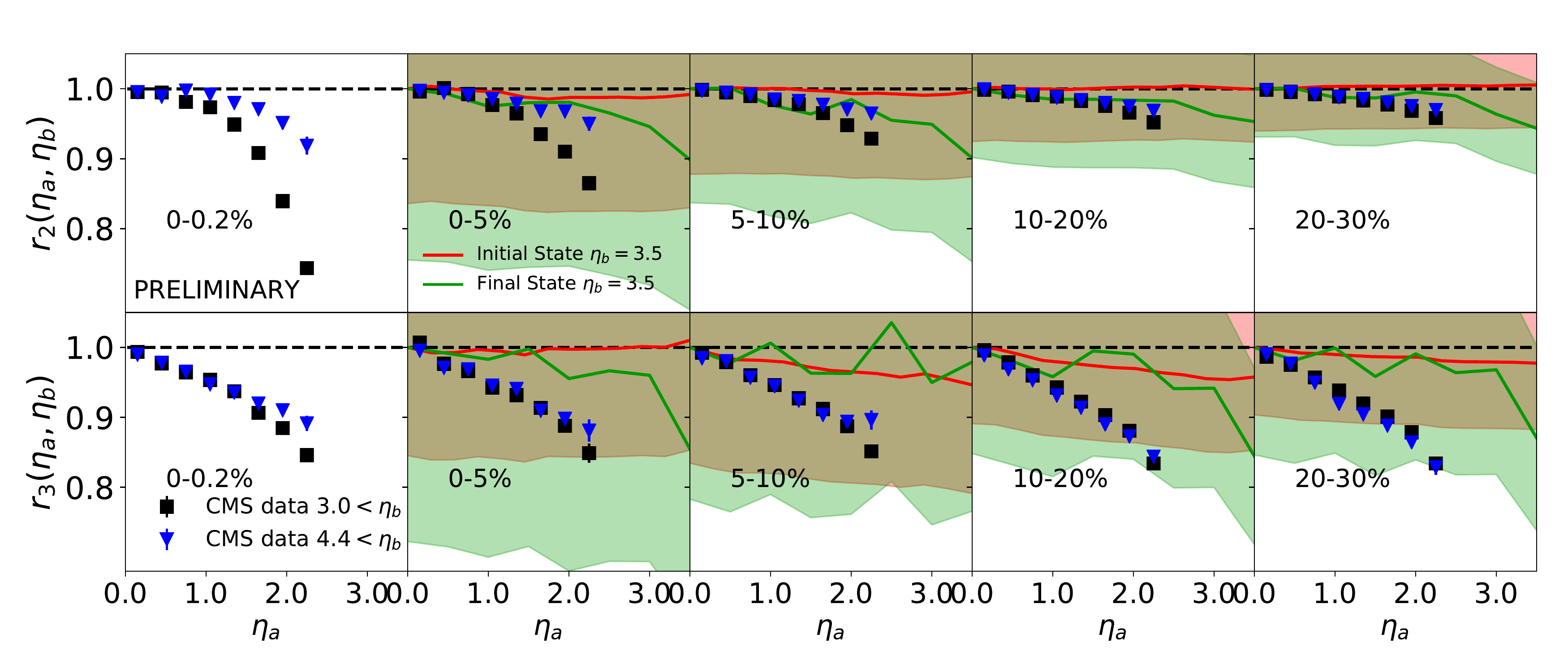}
    \caption{The quantity $\tilde{r}_n(\eta_a, \eta_b)$ labeled as ``initial state" and plotted in red, and  $r_n(\eta_a, \eta_b)$ labeled as ``final state" and plotted in green, as compared to CMS data from \cite{Khachatryan:2015oea}.}
    \label{fig:rn}
\end{figure}

From Fig. (\ref{fig:rn}), it is clear that relatively small initial state decorrelation leads to larger final state decorrelation. The $\varepsilon_n$'s only take information about the energy density but not the other components of $T^{\mu\nu}$, whereas the hydrodynamic evolution includes all components. Thus differences in these other components between rapidity slices could lead to larger decorrelation in the $v_n$ observable after hydro evolution, compared to that of the initial state quantity. Altering the rapidity profile according to Eq. (\ref{eq:eta_profile}), particle sampling, and hadronic evolution are all additional sources of decorrelation that affect the final state quantity $r_n(\eta_a, \eta_b)$, but not $\tilde{r}_n(\eta_a, \eta_b)$.  
\section{Conclusion}
The phenomenological effects of the initial state fluctuations introduced by JIMWLK on final state observables, including $v_2(\eta)$ and $r_n(\eta_a, \eta_b)$ were explored for the first time. Relatively small decorrelation of the initial state $\epsilon_n(\eta)$'s leads to larger final state decorrelation. With current statistics, the model underpredicts the decorrelation in the $r_n(\eta_a, \eta_b)$, leaving room for other sources of decorrelation such as thermal fluctuations in hydrodynamics, and minijets. 
\section{Acknowledgements}
This work is supported in part by the Natural Sciences and Engineering Research Council of Canada. Computation for this work was done on the Compute Canada supercomputer Beluga, maintained by Calcul Qu\'ebec. SM acknowledges funding from The Fonds de recherche du Qu\'ebec - Nature et technologies (FRQNT) through the Programme de Bourses d'Excellence pour \'Etudiants \'Etrangers (PBEEE).



%% References


\bibliographystyle{elsarticle-num}
\bibliography{Bibliography}






% \end{thebibliography}

\end{document}